# Simulated Performance Of Algorithms For The Localization Of Radioactive Sources From A Position Sensitive Radiation Detecting System (COCAE)

K. Karafasoulis[a,b], K. Zachariadou[a,c], S. Seferlis[a], I. Kaissas[a], C. Lambropoulos[d], D. Loukas[e], C. Potiriadis[a]

[a]*Greek Atomic Energy Commission, Patriarxou Grigoriou & Neapoleos,15310, Athens, Greece*
[b]*Hellenic Army Academy, 16673 Vari, Greece*
[c]*Technological Educational Institute of Piraeus, Thivon 250, 12244, Egaleo, Greece*
[d]*Technological Educational Institute of Chalkida, Psachna Evias, 34400 Greece*
[e]*Institute of Nuclear Physics, National Center for Scientific Research Demokritos 15310,Athens, Greece*

**Abstract.** Simulation studies are presented regarding the performance of algorithms that localize point-like radioactive sources detected by a position sensitive portable radiation instrument (COCAE). The source direction is estimated by using the List Mode Maximum Likelihood Expectation Maximization (LM-ML-EM) imaging algorithm. Furthermore, the source-to-detector distance is evaluated by three different algorithms based on the photo-peak count information of each detecting layer, on the quality of the reconstructed source image as well as on the triangulation method. These algorithms have been tested on a large number of simulated photons in a wide energy range (from 200keV up to 2MeV) emitted by point-like radioactive sources located at different orientation and source-to-detector distances.

## INTRODUCTION

The COCAE instrument [1] is a portable detecting system under development aimed to be used to accurately detect the position as well as the energy of radioactive sources in a broad energy range up to ~2 MeV, by exploiting the Compton scattering imaging technique [2]. Among its applications could be the security inspections at the borders and the detection of radioactive sources into scrap metals at recycling factories. The instrument consists of ten parallel planar layers made of pixelated Cadmium Telluride (CdTe) crystals occupying an area of 4cmx4cm, placed 2cm apart from each other. Each detecting layer consists of a two-dimensional array of pixels (100x100) of 400μm pitch, bump-bonded on a two-dimensional array of silicon readout CMOS circuits. Both pixels and readout arrays are on top of an $Al_2O_3$ supporting printed circuit board layer.

The Monte Carlo simulation studies regarding the development of the COCAE instrument encompass the following steps:

A) The exact COCAE detector geometry is modelled by an open-source object-oriented software library (MEGAlib [3]) which provides an interface to the Geant4 [4] toolkit that simulates the passage of particles through matter. Especially, the Compton scattering is accurately modelled by taking into account the influence of the Doppler broadening effect. Furthermore, since Compton interactions can occur not only in sensitive but also in non-sensitive materials of the detector resulting to incomplete energy measurements, special care has been taken to incorporate an accurate geometric and physical description of the detector's passive materials.
B) Point-like gamma sources with isotropic emission placed at different positions with respect to the detector are modeled emitting a large number of photons (~2x$10^9$) in an energy range varying from 60keV up to 2000keV.
C) The energy deposition ("hit") of each photon on the detector is recorded. The accurate simulation of the detector response is ensured by taking into account all relevant physical processes (Compton scattering, photoelectric effect, pair production, electron/positron transportation into matter, ionization). During the event reconstruction, dedicated algorithms are used in order to form events by using the spatial and energy information of each individual hit.

Important performance parameters of the COCAE instrument such as its detecting efficiency and angular resolution have been studied by Monte Carlo simulation [5]. Since the determination of the correct sequence of the photons interactions with the detector via the process of Compton scattering strongly affects the detector's efficiency of evaluating the direction of the incident photons, various techniques [6] (that exploit the kinematical and geometrical information of Compton scattering events as well as statistical criteria), have been extensively studied in order to select the best performing one [7]. Furthermore, the overall efficiency of event reconstruction has been evaluated in a wide range of initial photon energies [8].

The task of the current work is to exploit the performance of different techniques of localizing point-like radioactive sources. The source localization task includes the source direction estimation as well as the source-to-detector distance estimation.

The algorithms have been extensively tested on a large number of Monte Carlo simulated gamma photons (~2x$10^9$) emitted by point-like sources in an energy range varying from 200keV up to 2MeV, located at various distances from the COCAE detector model.

## DIRECTION ESTIMATION

Point-like radioactive sources have been reconstructed by applying the List Mode Maximum Likelihood Expectation Maximization (LM-MLEM) imaging algorithm [9]. According to this algorithm the image of a point source is generated by projecting each Compton event cone into an imaging plane and then by performing successive iterations on the back-projected image in order to find the source distribution with the highest likelihood of having produced the observed data.

Figure 1 depicts the reconstructed image in spherical coordinates of an 800keV point-like source, placed at $\theta = 26.56^0, \phi = 0^0$ by using the LM-MLEM imaging algorithm (50 iterations).

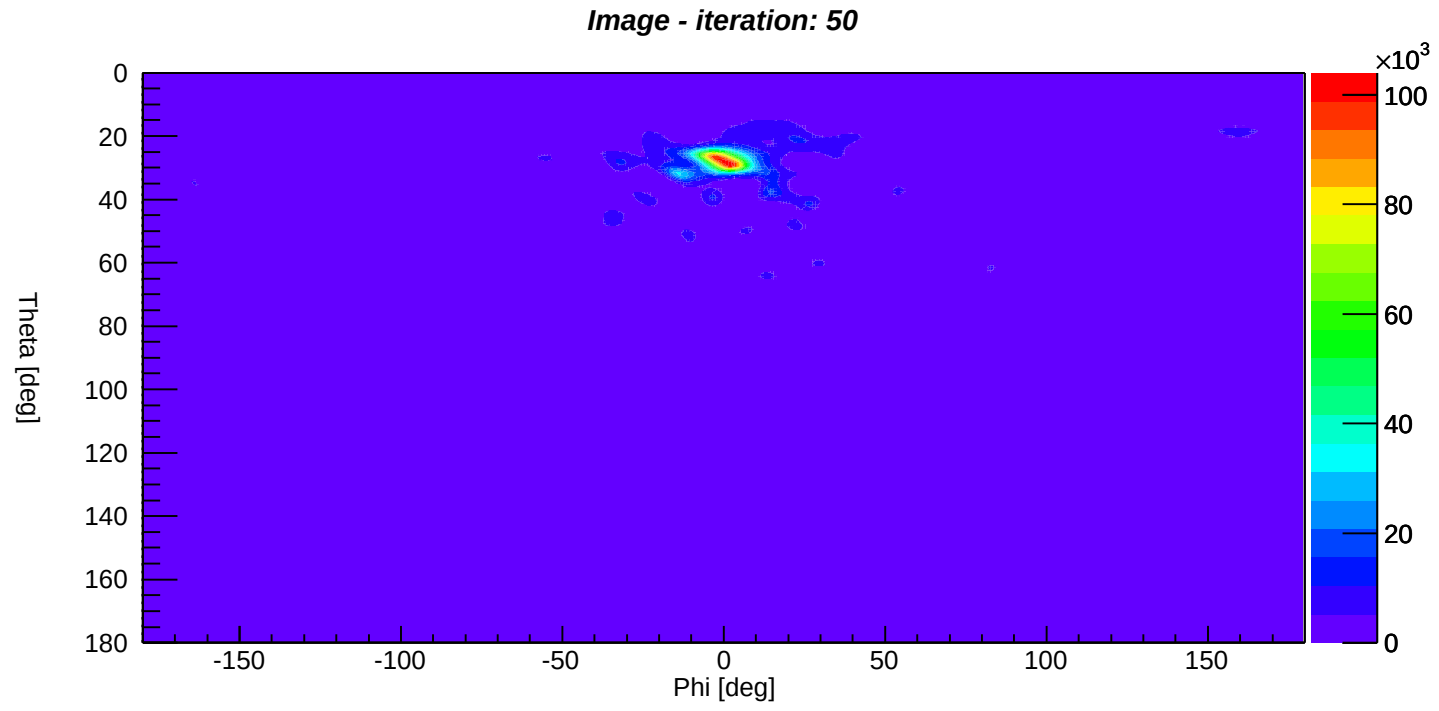

**FIGURE 1.** Reconstructed image of an 800keV point source located at $\theta = 26.56^0$, $\phi = 0^0$ from the detector's center, using the LM-MLEM image reconstruction algorithm.

The direction of point-like source has been estimated by the azimuth and the inclination distribution of the reconstructed image. Presented in Fig. 2a and Fig. 2b is the azimuth and the inclination distribution respectively for the case of an 800keV point-like source located at $\theta = 26.56^0$, $\phi = 0^0$.

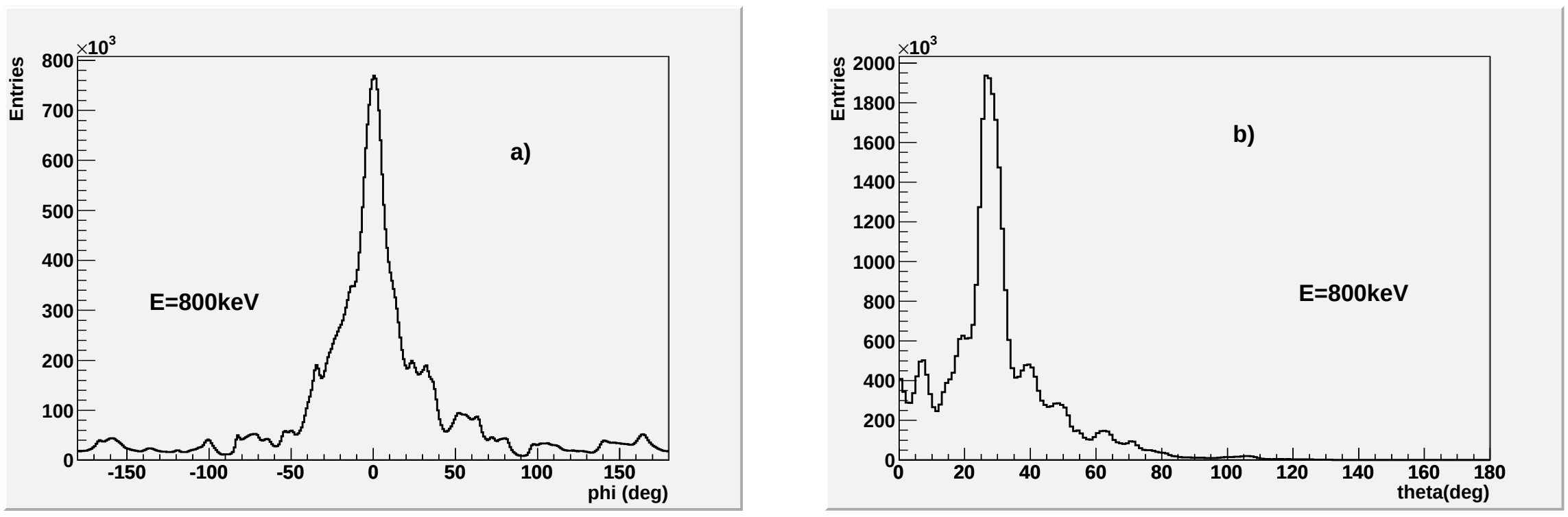

**FIGURE 2.** a) azimuth and b) inclination distributions of the reconstructed image of a 800keV point-like source located at $\theta = 26.56^0$, $\phi = 0^0$.

Shown in Fig. 3a and Fig. 3b is the reconstructed azimuth and inclination respectively as a function of the incident photon energy for the case of a point-like source located at $\theta = 44.84^0$ and $\phi = 0^0$ whereas Fig. 4 depicts the reconstructed inclination as a function of the real inclination for point-like sources with energy varying from 200keV up to 2MeV. It can be noticed that the azimuth as well as the inclination source's coordinate is estimated within less than one degree. Similar results have been obtained for inclination angles up to 50 degrees.

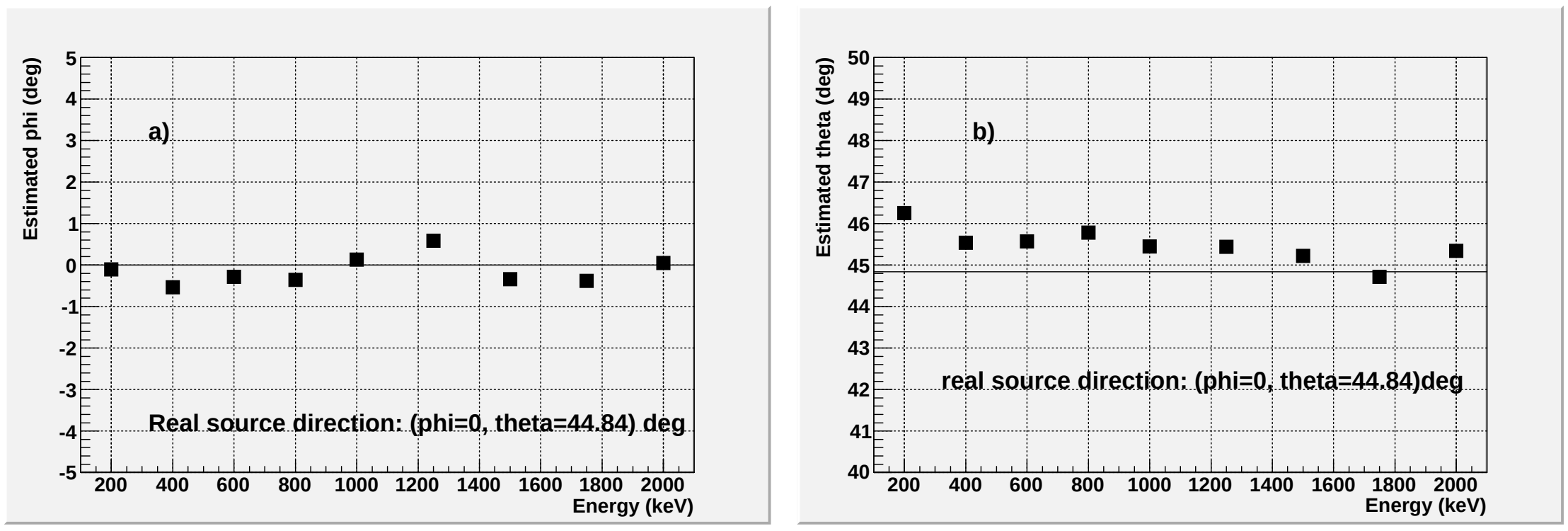

**FIGURE 3.** Estimated a) azimuth and b) inclination as a function of the incident photon energy for a point-like source located at $\theta = 44.84^0$, $\phi = 0^0$.

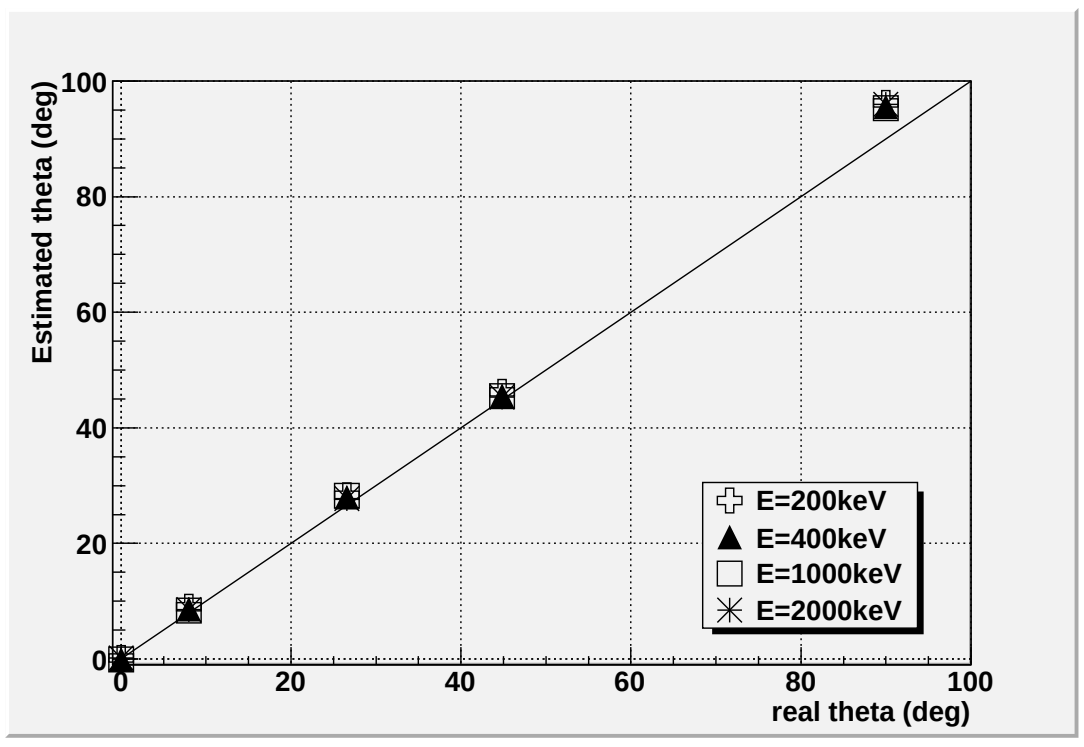

**FIGURE 4.** Estimated inclination as a function of the real inclination, for point-like sources emitting 200keV, 400keV, 1000keV and 2000keV photons.

# SOURCE-TO-DETECTOR DISTANCE ESTIMATION

Presented in the following sections is the evaluated performance of three different algorithms of estimating the distance of point-like radioactive sources based on a) the photo-peak count information from each detecting layer b) the quality of the reconstructed source image and c) the triangulation method.

## The Photo-Peak Count Information Technique

According to this method the estimation of the source-to-detector distance is based on the number of the fully absorbed photons (via a photoelectric effect) in each detecting layer [10].

For point-like sources located on the detector's axis of symmetry, the distance (d) of the source from the detector's first layer is evaluated by fitting the following function:

$$N_i \propto \exp\left( -(i-1)\left( \sum_j \mu_j t_j \right) \right) \cdot \frac{\sin^{-1}\left( \frac{k^2}{(d+(i-1)g)^2 + k^2} \right)}{\sin^{-1}\left( \frac{k^2}{d^2 + k^2} \right)} \tag{1}$$

where $N_i$ is the number of photo-peak counts recorded in each detecting layer (i), $t_j$ is the thickness of the $j^{th}$ material of each detecting layer with corresponding total absorption coefficient $\mu_j$, g is the distance between the layers and k is half the length of the detecting rectangular layer edge. The sum runs over all materials of the $i^{th}$ detecting layer. The first term of the above equation reflects the absorption by the front layers of the detector whereas the second term is the ratio of the solid angle of the $i^{th}$ detecting layer over the solid angle of the first detecting layer. The factor $a(E) = \sum_j \mu_j t_j$ in the absorption part is a function of the incident photon energy (E) and is evaluated for each incident photon energy E by simulating a point source placed on the detector's axis of symmetry at d=71 cm from its first detecting layer.

Figure 5a illustrates the estimated source-to-detector distance as a function of the real source-to-detector-distance, for different incident gamma rays energy whereas Fig. 5b depicts the estimated distance as a function of the incident gamma ray energy for a given source-to-detector distance (d=71.35cm).

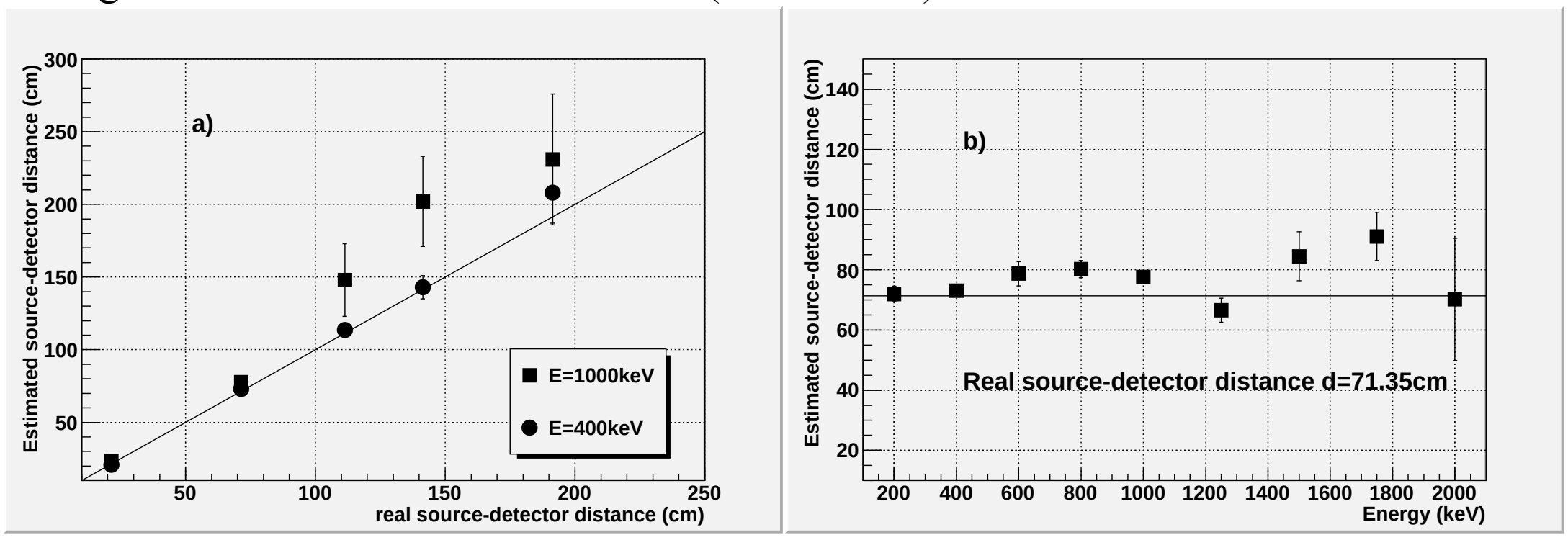

**FIGURE 5.** Estimated source-to-detector distance a) as a function of the real source-to-detector distance and b) as a function of the initial photon energy.

Results demonstrate that this method can estimate within 2σ source-to-detector distances up to 2m, for incident photon energies up to 2MeV. Higher statistics is necessary in order to reduce the errors for the case of incident photon energies above 1000keV emitted by sources located at distances greater than ~1m.

It has to be noticed that the distance estimation of point-like sources located off the detector's symmetry axis is performed by evaluating the source direction according to the method described in the previous section and then by aligning the detector with the source, so Eq. 1 can be applied.

## The Reconstructed Image Technique

The reconstructed image technique exploits the quality of the reconstructed image of point-like sources as a tool for estimating the source-to-detector distance [11]. It is based on the assumption that the quality of the image should be better (the FWHM of the x- and the y-distribution of the image has the lowest value) when the projection imaging plane is placed on the real source-to-detector-distance rather than in other distances. For studying the performance of this method a large number of photons emitted by point-like sources of different energies, located on the detector's axis of symmetry (x=y=0) and at various source-to-detector distances (z=20cm, z=30cm, z=50cm) has been simulated. Then the image of each point-like source has been reconstructed by the LM-MLEM imaging algorithm (200 iterations) at various projection imaging planes and the combined FWHM of the x- and the y- coordinate distributions is evaluated.

Shown in Fig. 6 is the combined FWHM for the case of a 400keV point-like source located on the detector's axis of symmetry and at various distances from the detector's center as a function of the projection imaging plane. Similar results have been obtained for point-like sources of incident photon energies up to 2000keV.

Results demonstrate that this method can only accurately estimate the distance of point sources being in the near field of the COCAE detector (distances up to z=30cm).

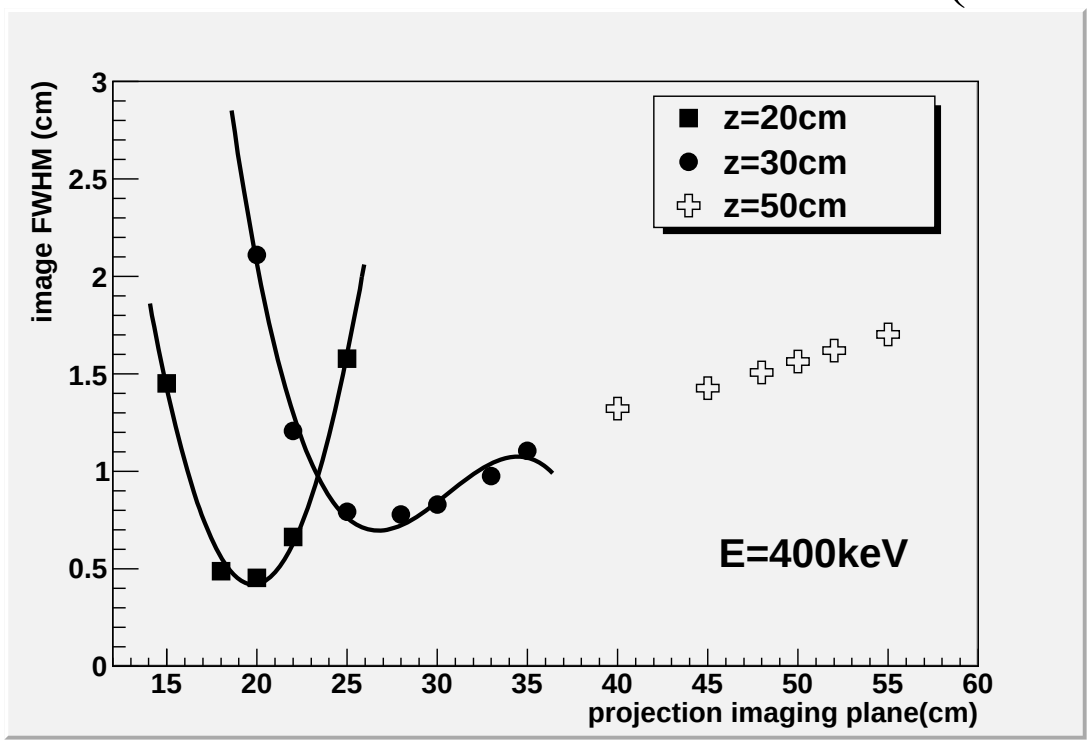

**FIGURE 6.** The FWHM of reconstructed images as a function of the projection imaging plane for 400keV photons emitted by point-like sources located at z=20cm, z=30cm and z=50cm from the detector's center.

## The Triangulation Technique

The ability of COCAE instrument to localize radioactive point-like sources by using the triangulation technique has also been exploited. To test the method, simulation data obtained by locating the COCAE instrument at two different positions at a given separation distance (d=7cm, d=25cm, d=50cm) have been used in order to identify the direction of the point-like source, located at $(r,\theta,\varphi)=(50cm,0^o,0^o)$. An algorithm estimates the position of the radioactive source as the middle of the minimum distance vector of the two 3D skew lines defined by the estimated source directions and the detector location.

Figure 7 depicts the estimated source position as a function of the incident photon energy for three separation distances of the COCAE models (d=7cm, d=25cm, d=50cm) when the source is located at $(r,\theta,\varphi)=(50cm,0^o,0^o)$.

It has to be pointed out that the minimum distance of the two 3D photon directions estimated by the two COCAE models is of the order of 3mm, reflecting the good performance of the method.

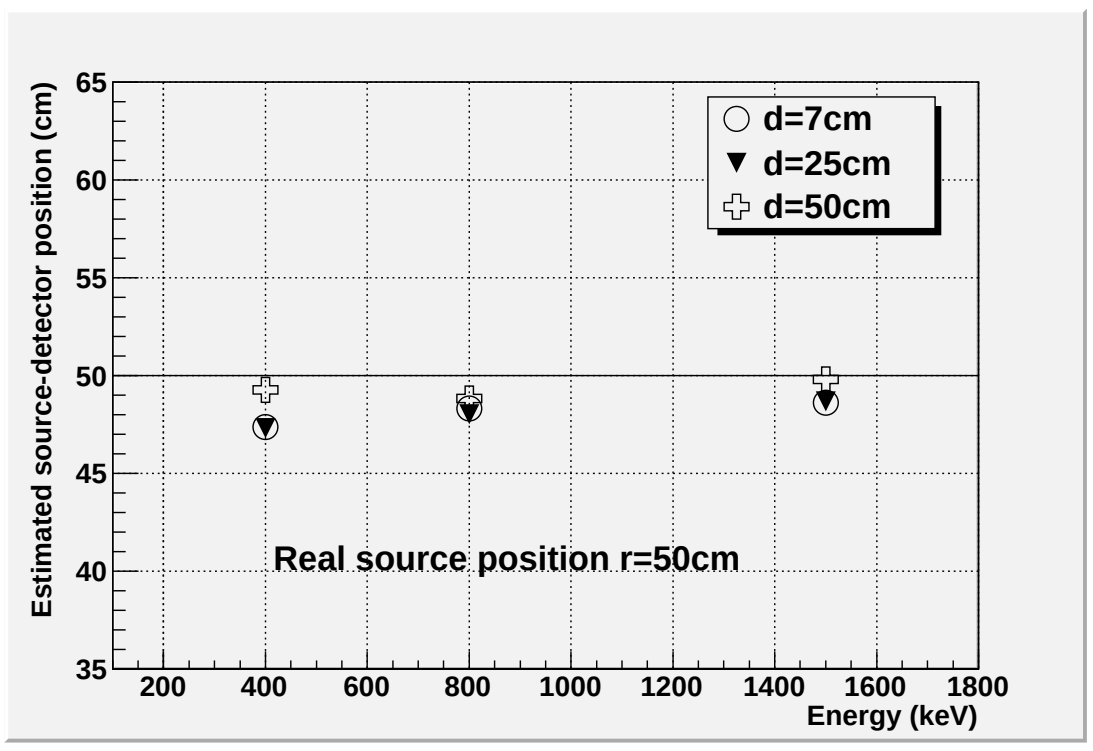

**FIGURE 7.** The triangulation method: Estimated source position as a function of the incident photon energy by locating the COCAE instrument at two different positions at a given separation distance (d=7cm, d=25cm, d=50cm). The source is located at $(r,\theta,\varphi)=(50cm,0^o,0^o)$.

# CONCLUDING REMARKS

Extensive simulations have been performed regarding the evaluation of the performance of algorithms that estimate the direction and the distance of point-like radioactive sources.

As far as the source direction estimation is concerned, simulation studies have demonstrated that it is accurately evaluated within one degree (both in azimuth and inclination) for inclination angles up to 50 degrees, by reconstructing the source's image using the LM-MLEM imaging algorithm. Further studies are underway for larger inclination angles and source-to-detector distances.

For the source-to-detector distance estimation, simulation studies have demonstrated that the algorithm based on the number of the fully absorbed photons in each detecting layer can estimate (within $2\sigma$) source-to-detector distances up to 2m, for incident photon energies up to 2MeV. Higher statistics is necessary in order to reduce the errors for the case of incident photon energies above 1000keV emitted by sources located at distances greater than ~1m.

An algorithm based on the quality of the reconstructed image of point-like sources for estimating source-to-detector distances has also been studied. Results demonstrate that this method can accurately estimate source-to-detector distances in the near field of the detector (up to ~30cm from its center).

The triangulation technique has also been exploited as a method of estimating the position of point-like sources. Results demonstrate that this method can estimate the position of point-like sources within few centimeters.

# ACKNOWLEDGMENTS

This work is supported by the European Community's Seventh Framework Program FP7-SEC-2007-01.